\documentclass{article}[11pt]

\usepackage{graphicx}
\usepackage{amssymb}
\usepackage{amsmath}
\usepackage{amsfonts}
\usepackage[font=small,labelfont=bf]{caption}
\usepackage{amsthm}
\usepackage{fullpage}
\usepackage{subfigure}

\newcommand{\frechet}{Fr\'echet}
\newcommand{\dfre}{d_F}

\newcommand{\dist}{\mathit{d}}
\newcommand{\D}{\mathcal{D}}
\newcommand{\BO}{\mathcal{O}}

\newcommand{\dfd}{discrete \frechet\ distance}
\newcommand{\npc}{\textbf{NP}-complete}

\newtheorem{definition}{Definition}
\newtheorem{theorem}{Theorem}

\begin{document}

\title{An Interesting Gadget for Chain Pair Simplification}

\author{Tim Wylie}

\date{}
\maketitle
\begin{center}
\vspace{-.5cm}
\small
Department of Computer Science \\ University of Texas - Rio Grande Valley \\ timothy.wylie@utrgv.edu
\end{center}
\begin{abstract}
In this paper we present an interesting gadget based on the chain pair simplification problem under the \dfd\ (CPS-3F), which allows the construction of arbitrarily long paths that must be chosen in the simplification of the two curves.  A pseudopolynomial time reduction from set partition is given as an example. For clarification, CPS-3F was recently shown to be in \textbf{P}, and the reduction is merely to show how the gadget works.
\end{abstract}
\section{Introduction}

The Chain Pair Simplification problem under the \dfd\ is an interesting problem dealing with how well two polygonal curves can be simplified in relation to each other with a minimum number of points retained.  Under the Hausdorff distance, the problem is \npc~\cite{Bereg:2008:SPC}. Recently, CPS-3F was shown to be in \textbf{P} with an algorithm that runs in $\BO(m^2n^2\min\{m,n\})$ where $m$ and $n$ are the number of nodes on the two polygonal chains \cite{Fan:2015:WADS}. 

Here, we exploit a trick from \cite{Wylie:2013:TCBB} where there are only two possible paths to simplify the curve, and we then modify this to allow for arbitrarily long sequences of simplification.  This can be thought of as a partition problem between the two chains, however, any actual mapping requires pseudopolynomial time.  The reason for this research was merely to see if this could be done.  It is not easy to create two curves with this property, but this document gives a method for building pairs of polygonal curves that can only be simplified in specified ways.  Further, the paths can be arbitrarily long before the curves come together again, and as many of these paths can be linked together as desired. This may be useful in proving other properties about the problem, such as in \cite{Wylie:2013:TCBB}.

For background information on the problems, the \dfd, etc., please refer to the references.

\section{Weighted Chain Pair Simplification} \label{sec:weighted}

We first cover a general version of CPS-3F introduced in \cite{Fan:2015:WADS}.  The weighted CPS problem allows for arbitrary weights on the vertices, and we want to find a solution such that neither of the two simplified chains have vertices whose weights sum to greater than a given value. The problem is formally defined as follows.

\begin{definition}[Weighted Chain Pair Simplification] \hfill \\
   \noindent \textbf{Instance:} Given a pair of 3D chains $A$ and $B$, with lengths $m$ and $n$,
    respectively, an integer $K$, three real numbers $\delta_1,\delta_2,\delta_3>0$,
    and a weight function $C:\{a_1,\dots, a_m,b_1,\dots, b_n\} \rightarrow \mathbb{R}^+$. \\
    \textbf{Problem:} Does there exist a pair of chains $A'$,$B'$ where $C(A'),C(B') \leq K$,
    such that the vertices of $A'$,$B'$ are from $A,B$ respectively, where  
    $\dfre(A,A') \leq \delta_1$, $\dfre(B,B') \leq \delta_2$, and $\dfre(A',B') \leq \delta_3$?

\end{definition}

\begin{theorem}
    The weighted chain pair simplification problem under the \dfd\ is weakly \npc \cite{Fan:2015:WADS}. 
\end{theorem}

\begin{figure}[t!]
    \centerline{\includegraphics[width=.2\textwidth]{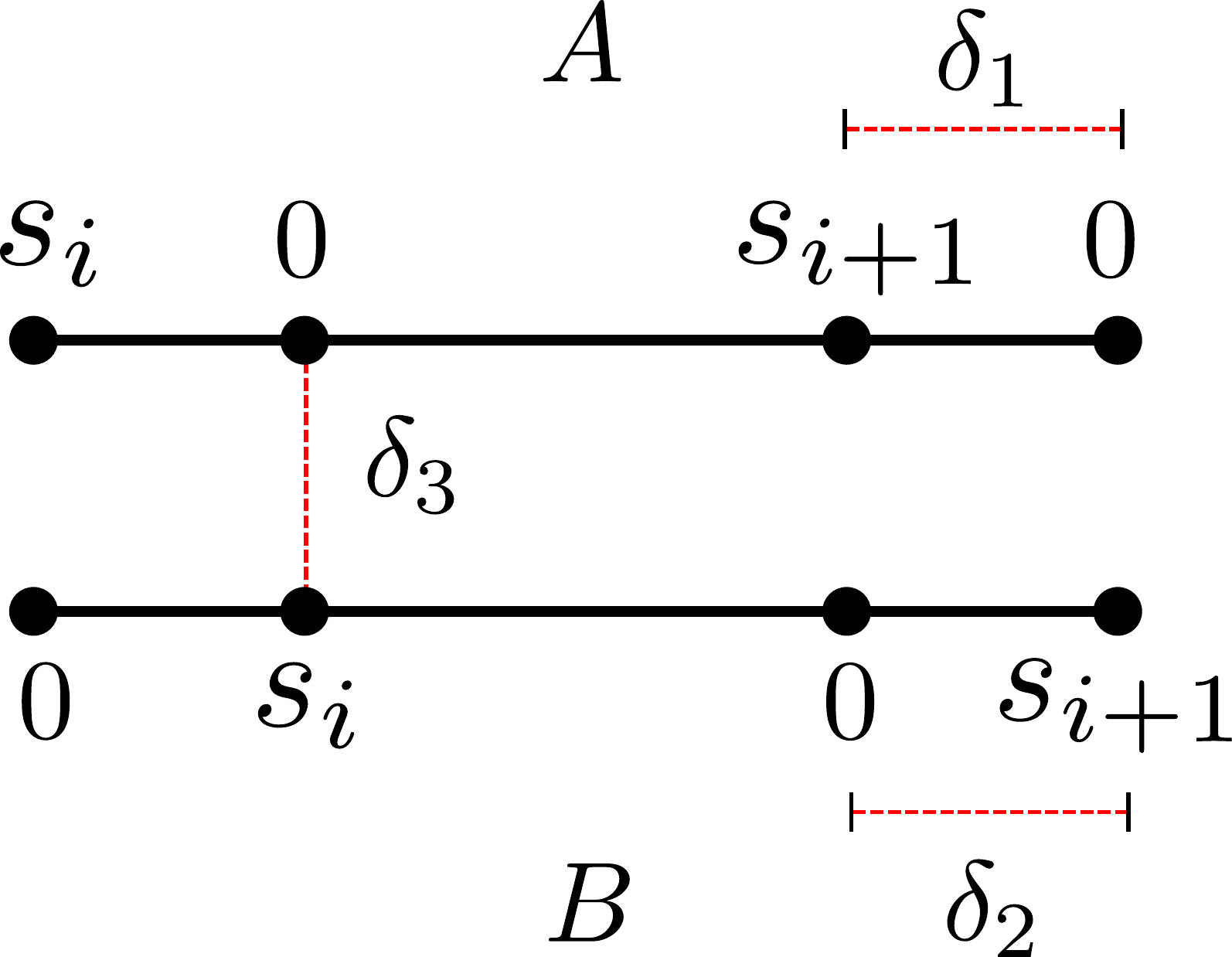}}
    \caption{The reduction for the weighted CPS-3F problem.}
    \label{fig:wcps}
\end{figure}  

The idea for this proof is a reduction from the Set Partition problem as seen in Figure \ref{fig:wcps}. Given a set of positive integers $S=\{s_1,\dots,s_n\}$, find two sets $P_1,P_2 \subset S$ such that $P_1 \cap P_2 = \emptyset$, $P_1 \cup P_2 = S$, and the sum of the numbers in $P_1$ equals the sum of the numbers in $P_2$. This is a weakly \npc\ special case of the classic subset-sum problem.

The reduction builds two curves with weights reflecting the values in $S$. We think of the two curves as the subsets of the partition of $S$. Although the problem requires positive weights, we also allow zero weights in our reduction for clarity. It is easy to modify the reduction to have positive weights and reside in a single dimension. Increasing all the weights by one gives $w(a_{2i-1})=w(b_{2i})=s_i+1$ and $w(a_{2i})=w(b_{2i-1})=1$. This adds one for every element of $S$ to both chains, so with $K=|S| + \mathfrak{S}/2$ the reduction is the same.


\section{Chain Pair Simplification (CPS-3F)} \label{sec:cps}

We now turn our attention to CPS-3F, which is the special case of the general weighted CPS problem where every vertex has a weight equal to one. The construction is similar to the weighted version but more involved. The formal definition of the problem is as follows.

\begin{definition}[Chain Pair Simplification]\hfill \\
    \textbf{Instance:} Given a pair of polygonal chains $A$ and $B$, with lengths $O(m),O(n)$
    respectively, an integer $K$, and three real numbers $\delta_1,\delta_2,\delta_3>0$.\\
    \textbf{Problem:} Does there exist a pair of chains $A'$,$B'$ each of at most $K$ vertices
    such that the vertices of $A'$,$B'$ are from $A,B$, respectively, where
    $\dfre(A,A') \leq \delta_1$,$\dfre(B,B') \leq \delta_2$, and $\dfre(A',B') \leq \delta_3$?
\end{definition}

In order to explain the reduction, it is first necessary to give some notation.
Given two polygonal chains $A=\langle a_1,\dots,a_m\rangle$, and $B=\langle b_1,\dots,b_n\rangle$, and constraints $\delta_1,\delta_2,\delta_3 \in \mathbb{R}^+$, 
we define rectangles which will be used for our construction. First, let $\D=\{(a_i,b_j) |$ $a_i \in A,$ $b_j \in B$ and $\dist(a_i,b_j) \leq \delta_3\}$.  This is the set of all pairs of nodes between the two chains that are at a distance of at most $\delta_3$ from each other.

\begin{figure}[h!]
    \centerline{\includegraphics[width=.58\textwidth]{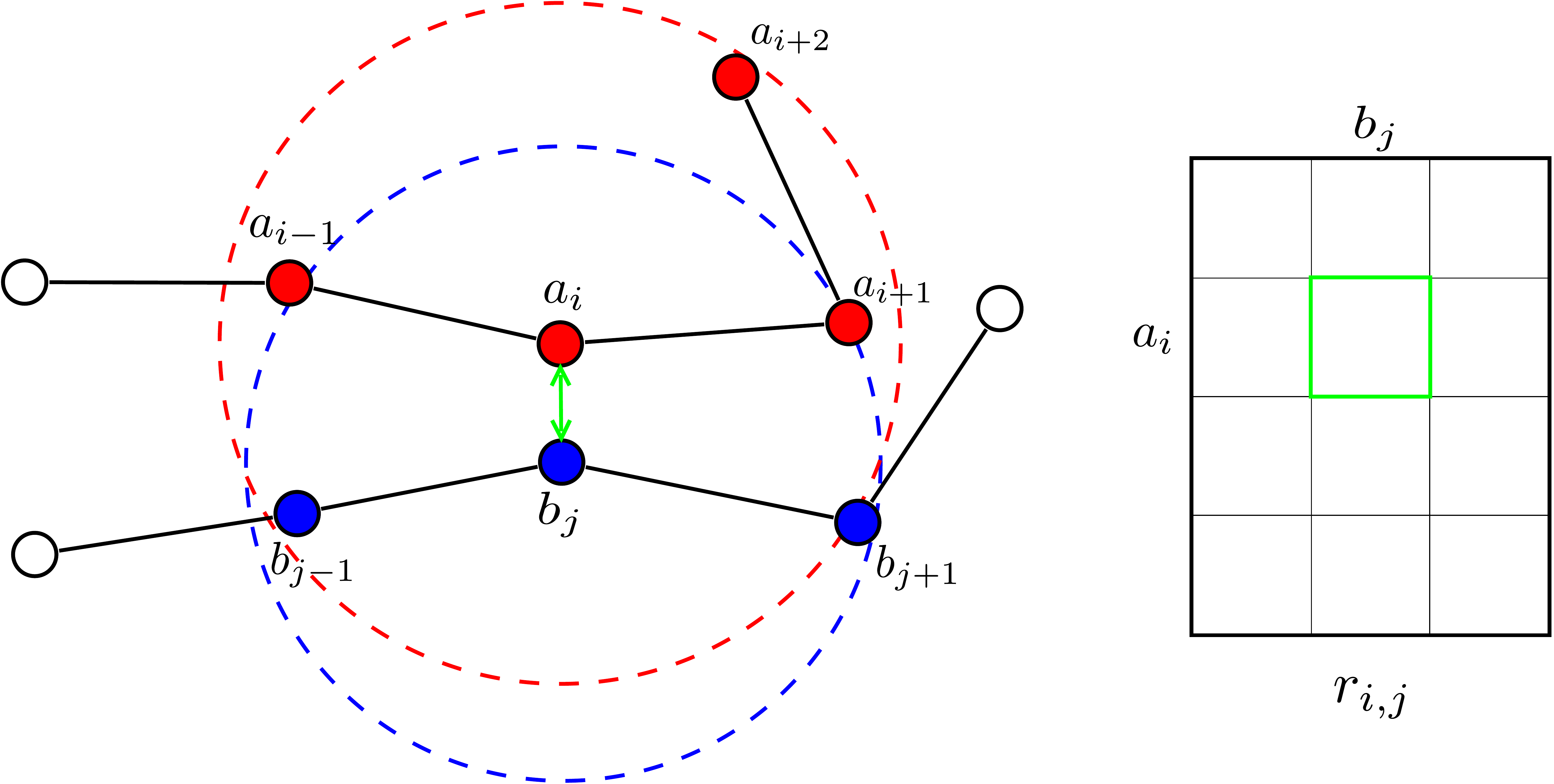}}
    \caption{The rectangle $r_{i,j}$ constructed 
        from subchains of $A,B$ where $\dist(a_i,b_j)\leq \delta_3$.  Here $S_A(a_i,\delta_1)$ contains 
        the vertices $a_{i-1}$ to $a_{i+2}$, and $S_B(b_j,\delta_2)$ contains
        the vertices $b_{j-1}$ to $b_{j+1}$.  Thus, $r_{i,j}$ is defined by the min and max node indices 
        in each subchain.}
    \label{fig:convert}
\end{figure}  

We define $S_X(x_i, \delta)$ as the maximal continuous subchain containing $x_i$ on the polygonal chain $X$ such that all the vertices on this subchain are contained in the sphere centered at $x_i$ and with radius $\delta$.  Let $\min$ and $\max$ refer to the minimum or maximum indexed element within $S_X(x_i,\delta)$. Now let $r_{i,j}$ be the rectangle defined as $\langle \min(S_A(a_i,\delta_1)),$ $\max(S_A(a_i,\delta_1)),$ $\min(S_B(b_j,\delta_2)),$ $\max(S_B(b_j,\delta_2)) \rangle $ such that $(a_i,b_j) \in \D$. A rectangle $r_{i,j}$ covers all the cells in a grid that are analogous to vertices in $S_A(a_i,\delta_1) \cup S_B(b_j,\delta_2)$ as shown in Figure \ref{fig:convert}.

\begin{figure}[t!]
    \begin{center}
        \subfigure[]{\label{fig:gadgeta}\includegraphics[width=.38\textwidth,keepaspectratio]{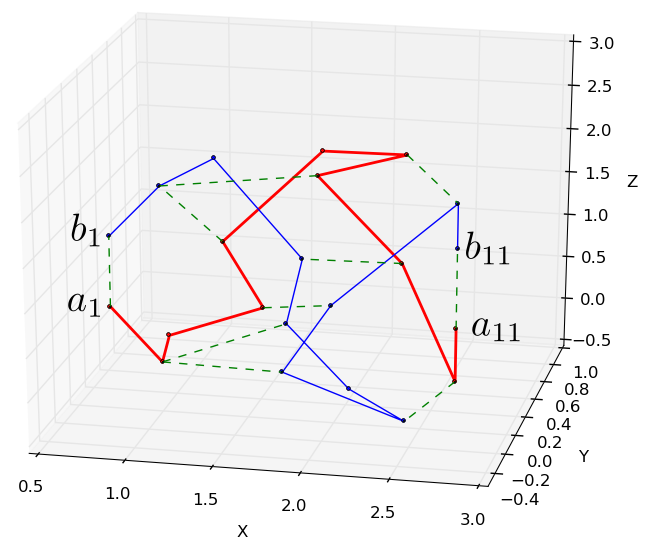}} 
        \hspace*{.5cm}
        \subfigure[]{\label{fig:gadgetrects}\includegraphics[width=.35\textwidth,keepaspectratio]{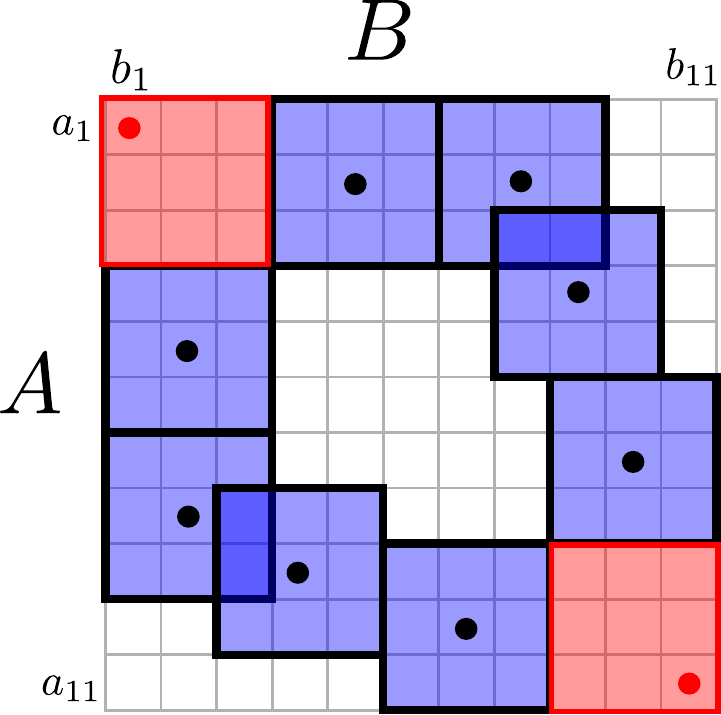}}  \\
    \end{center}
    \caption{(a) The chains listed in Table \ref{tab:nodes}. The dotted lines connect pairs of vertices $(a_i,b_j)$ between the two
        chains where $\dist(a_i,b_j) \leq \delta_3$. (b) The relationship in rectangles with the dots representing
        the pairs between the chains where $\dist(a_i,b_j) \leq \delta_3 = 0.9$ which create the rectangles when $\delta_1=\delta_2=1.1$.}
    \label{fig:cps_gadget}
\end{figure}

Marking just the pairs of nodes in the grid $A \times B$ that are within $\delta_3$ of each other (the pairs in $\D$) is equivalent to a discrete version of the often-used free-space diagram for the continuous \frechet\ distance \cite{Alt:1995:JCOMPS}. The discrete version was similarly used in \cite{Avraham:2013:CORR}.  Here, we have extended this idea to incorporate the possibility of simplifying $A$ and $B$ as well.

\begin{table}[h]
\renewcommand{\arraystretch}{1.0} 
\centering
  \begin{tabular}{|c | c | c | c |}
    \hline
    $A$ & ($x, y, z$) & $B$ & ($x, y, z$) \\ \hline
    $a_1$   &   (0.70, 0.00, 0.60)    & $b_1$      & (0.70, 0.00, 1.40)  \\ 
    $a_2$   &   (1.00, 0.00, 0.00)    & $b_2$      & (1.00, 0.00, 2.00)  \\
    $a_3$   &   (1.28, $-$0.56, 0.90) & $b_3$      & (0.96, 0.88, 1.55)  \\ 
    $a_4$   &   (1.78, $-$0.49, 1.20) & $b_4$      & (1.64, 0.49, 0.80)  \\
    $a_5$   &   (1.60, $-$0.60, 2.00) & $b_5$      & (1.57, 0.42, 0.10)  \\ 
    $a_6$   &   (2.06, $-$0.35, 2.80) & $b_6$      & (1.92, 0.49, $-$0.70)  \\
    $a_7$   &   (2.41, 0.00, 2.50)    & $b_7$      & (2.41, 0.00, $-$0.50)  \\ 
    $a_8$   &   (1.74, 0.46, 1.80)    & $b_8$      & (1.78, $-$0.21, 0.20)  \\
    $a_9$   &   (2.24, 0.46, 1.85)    & $b_9$      & (2.13, $-$0.42, 1.20)  \\ 
    $a_{10}$&   (2.70, 0.00, 0.00)    & $b_{10}$   & (2.70, 0.00, 2.00)  \\
    $a_{11}$&   (2.70, 0.00, 0.60)    & $b_{11}$   & (2.70, 0.00, 1.50)  \\ 
    \hline
  \end{tabular}
  \caption{The nodes in the chains for $A$ and $B$.}
  \label{tab:nodes}
\end{table}

Similar to WCPS-3F, we show a reduction from the set partition problem, but note that the reduction is pseudopolynomial, and exponential in the size of the input. Given the set $S=\{s_1,s_2,\dots, s_n\}$ where $s_i \in \mathbb{Z}^+$ $\forall$ $0 \leq i \leq n$, we want to find two sets $P_1 \subset S,P_2 \subset S$ such that $P_1 \cap P_2 = \emptyset$, $P_1 \cup P_2 = S$, and $\sum^{|P_1|}_{j=1}P_{1_j} = \sum^{|P_2|}_{j=1}P_{2_j}$.

The basis of the reduction is the relationship between the two chains listed in Table \ref{tab:nodes} and shown in Figure \ref{fig:gadgeta}. In the figure the two curves are shown with dotted lines between the pairs of vertices within $\delta_3=0.9$ of each other. Based on our definition for rectangles, these chains correspond to the ten rectangles shown in Figure \ref{fig:gadgetrects}. Based on $\delta_1 = \delta_2 = 1.1$ and $\delta_3 = 0.9$, the pairs of nodes between the $A,B$ are $(a_1, b_1)$, $(a_2, b_5)$, $(a_2, b_8)$, $(a_4, b_9)$,  $(a_5, b_2)$, $(a_7, b_{10})$, $(a_8, b_2)$, $(a_9, b_4)$, $(a_{10}, b_7)$, and $(a_{11}, b_{11})$. Every node $h_i$ on either chain is also within $\delta_1=\delta_2=$1.1 of both its neighbors ($x_{i-1}, x_{i+1}$), but the distance is greater than 1.1 for the other nodes on the chain.  The only exception is the first and last node which is within 1.1 of two nodes. Note that if $(a_1,b_1)$ and $(a_{11},b_{11})$ are moved such that they are not within $\delta_3$ of each other, then the configuration in Figure \ref{fig:gadgetrects} would be missing the top left and bottom right rectangles. These are only needed when beginning or ending a chain. This will be clear in the reduction.

The configuration in Figure \ref{fig:cps_gadget} works by having only two possible simplifications. Ignoring the starting and ending squares, one path yields chains (for $A',B'$) of lengths (3,4) while the other yields lengths of (4,3). We can think of this as both having 3 added to them, and then choosing which chain to add the extra node to. Now, if we stack them along the diagonal (again without the first and last rectangles). This would mean that the two gadgets would connect such that the only two paths would give simplified chains of (8,6) or (6,8). In this way, we can create chains such that only two possible simplifications exist with any chosen difference between the two.

In order for this to work, we need another chain to add on to the one defined in Table \ref{tab:nodes} such that the last three nodes in each chain are replaced by the first three nodes in the new chain, but the chain must also maintain all the relationships. Chains $A$ and $B$ were designed so that the beginning pair ($a_2, b_2$) align with the end pair ($a_{10}, b_{10}$) and each chain can be copied and ``attached'' to the end of the previous chain by a translation in the positive $x$ direction.  By combining the new chains, $a_2$ replaces the node as $a_{10}$ and equivalently for $B$.

Let $H_{i\dots j}$ represent the consecutive subchain of a polygonal curve $H$ from node $h_i$ through $h_j$. 
Then let $T^{(j,k)}$ denote the sequence $T$ translated by $1.7j$ units in the positive \textbf{X} direction and $10k$ in the positive \textbf{Z} direction, e.g., for all $t_i \in T^{(2,3)}$ the coordinates are $t_i = (x_i + 3.40,$ $y_i,$ $z_i + 30)$. The value $1.70$ is the distance between $a_2$ and $a_{10}$ and between $b_2$ and $b_{10}$, which are both aligned along the \textbf{X}-axis.

\begin{figure}[t!]
    \begin{center}
        \subfigure[]{\label{fig:gadget3a}\includegraphics[width=.5\textwidth,keepaspectratio]{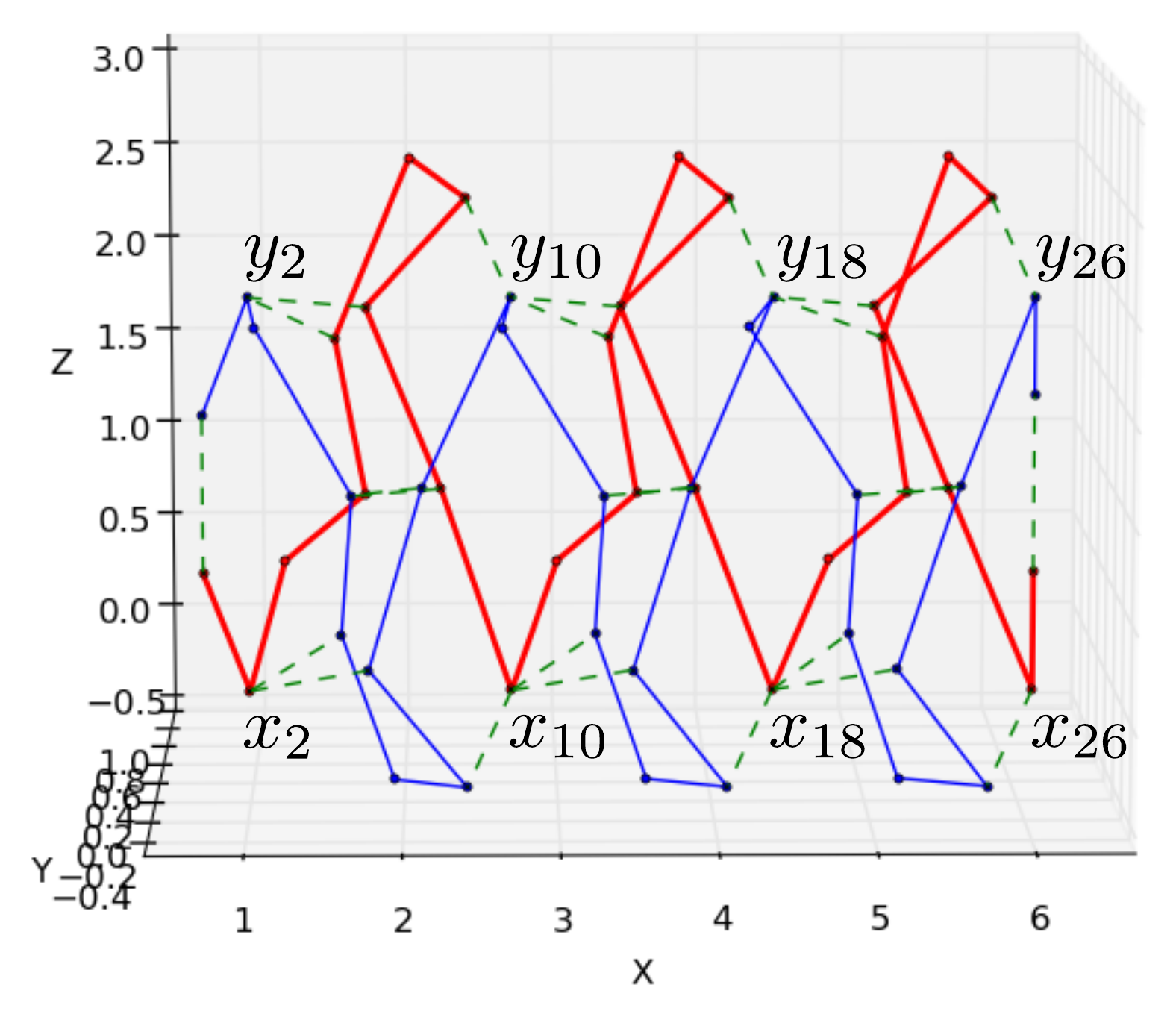}} 
        \subfigure[]{\label{fig:gadget3b}\includegraphics[width=.47\textwidth,keepaspectratio]{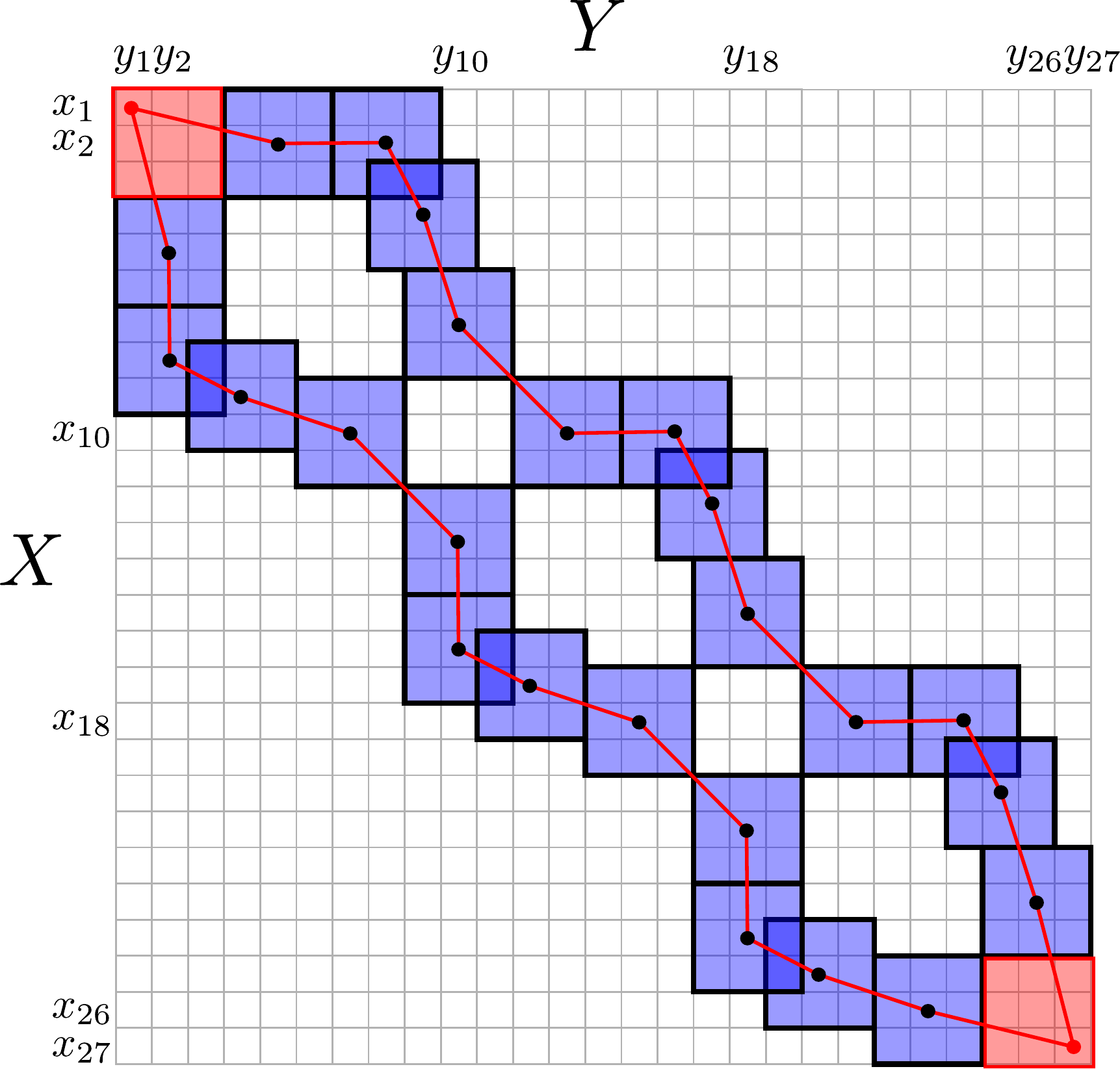}}  \\
    \end{center}
    \caption{(a) A sequence of three repeated gadgets with the beginning and end nodes. 
        (b) The resulting rectangles from figure (a). There are only two paths
        since the paths are monotonically non-decreasing. The possible simplified
        chains are of lengths (9,12) or (12,9). This is equivalent to $s_i=3$ with 9 extra vertices 
        in each chain.}
    \label{fig:cps_gadget3}
\end{figure}

Now constructing a new chain to represent $s_i=3$, we denote it as $X = \langle A^{(1,0)}_{1\dots 9},$ $A^{(2,0)}_{2\dots 9},$ $A^{(3,0)}_{2\dots 11} \rangle$, and $Y = \langle B^{(1,0)}_{1\dots 9},$ $B^{(2,0)}_{2\dots 9},$ $B^{(3,0)}_{2\dots 11} \rangle$.
The two polygonal chains are shown in Figure \ref{fig:gadget3a}, and the rectangle representation of their relationship is shown in Figure \ref{fig:gadget3b}. The only possible simplifications are highlighted in \ref{fig:gadget3b}, and they yield chains ($X',Y'$) of lengths (9,12) or (12,9).




    
    The general reduction algorithm is as follows. 
    Given $S=\{s_1,s_2,\dots, s_n\}$ from the set partition problem. 
    For all $s_i$ where $s_i \in S$, create 
    $X_{s_i}=\langle A^{(1,i)}_{1}$, $A^{(1,i)}_{2\dots 9}$, $A^{(2,i)}_{2\dots 9}, \dots, A^{(s_{i}-1,i)}_{2\dots 9}, A^{(s_i,i)}_{2\dots 9}$, $A^{(s_i,i)}_{10}, A^{(s_i,i)}_{11}\rangle$. 
    You can think of each sequence from two to nine as a ``one''. 
    Thus, if $s_i=5$, you will need the starting node, five sequences from two to nine, and then the ending nodes. 
    Since ten and two represent the same node, ten is omitted except at the end when no more sequences are added.
    We also create a $Y_{s_i}$ in the same manner using subsequences of $B$.
    Finally, let $X=\langle X_{s_1}, \dots, X_{s_n} \rangle$ and $Y=\langle Y_{s_1}, \dots, Y_{s_n} \rangle$.
    
    Now we can state that for an instance of the set partition problem with set $S$, $S$ can be evenly partitioned if and only if $X$ and $Y$
    can be simplified via CPS-3F such that $\delta_1=\delta_2=1.1$, $\delta_3=0.9$, and $K= 2|S| + 3\mathfrak{S} + \mathfrak{S}/2$.
    Here, $\mathfrak{S}$ is the sum of the elements of $S$.  Note that since we create $t$ gadgets for each number $t$, the reduction is polynomial in the numeric value of the input, and exponential in the length of the input numbers to the set partition problem.

\section{Conclusion} \label{sec:conclusion}

In this paper we covered an interesting construction of two polygons as input for the CPS-3F problem and a pseudopolynomial time reduction from set partition to CPS-3F as an example of its use.  This gadget may prove useful in constructing specific arrangements for other applications or may simply be an enjoyable excursion into part of the character of the chain pair simplification problem.



\nocite{Wylie:2013:TCBB}
\nocite{Fan:2015:WADS}
\bibliography{proof}
\bibliographystyle{abbrv}

\end{document}